\documentclass[aps,prb,preprint,floatfix]{revtex4-1}

\usepackage{graphicx}
\usepackage{amsmath}
\usepackage{amsfonts}
\usepackage{amssymb}
\usepackage{dcolumn}
\usepackage{physics}
\usepackage{commath}
\usepackage{epstopdf}
\usepackage{hyperref}
\usepackage[utf8]{inputenc}
\usepackage[normalem]{ulem}

\usepackage{stackengine}
\newcommand{\RNum}[1]{\uppercase\expandafter{\romannumeral #1\relax}}

\graphicspath{{pics/}}

\renewcommand\bra[1]{{\langle{#1}|}}
\renewcommand\ket[1]{{|{#1}\rangle}}

\usepackage{color}
\definecolor{darkgreen}{RGB}{38, 175, 79}

\begin{document}

\title{A quantum moat barrier, realized with a finite square well}

\author{A. Ibrahim$^{1}$}

\author{F.~Marsiglio$^{2}$}
\email{fm3@ualberta.ca}

\affiliation{$^{1}$Department of Physics and Astronomy, University of Waterloo, Waterloo, Ontario, Canada, N2L~3G1}
\affiliation{$^{2}$Department of Physics, University of Alberta, Edmonton, Alberta, Canada, T6G~2E1}

\begin{abstract}

The notion of a double well potential typically involves two regions of space separated by a repulsive potential barrier. The solution is a wave function that is suppressed in the barrier region and localized in the two surrounding regions. Remarkably, we illustrate that similar solutions can be achieved using an attractive ``barrier'' potential (a ``quantum moat'') instead of a repulsive one (a ``quantum wall''). The reason this works is intimately connected to the concepts of ``orthogonalized plane waves'' and the pseudopotential method, both originally used to understand electronic band structures in solids. While the main goal of this work is to use a simple model to demonstrate the barrier-like attribute of a quantum moat, we also show how the pseudopotential method is used to greatly improve the efficiency of constructing wave functions for this system using matrix diagonalization.

\end{abstract}

\date{\today }
\maketitle

\section{Introduction}

There are many fascinating aspects of quantum mechanics which clearly disturb out classical intuition. Most models of tunneling begin by introducing two ``free'' regions of space separated from one another by some kind of barrier.\cite{doublewell,jelic12,dauphinee15} The solution that describes the tunneling particle oscillates in the two free regions and rapidly decays in the barrier region. As a result, the particle's wave functions are ``cat-like,'' i.e. the particle co-exists in the two free regions, with some non-zero probability of tunneling through the barrier region. 

In most demonstrations of tunneling, the model barrier is a positive, i.e. repulsive, potential wall. However, similar behaviour arises even when the ``barrier'' is a negative, i.e. attractive, potential.\cite{ibrahim18} As an analogy, it is as if the particle, instead of trying to pass through a wall, is struggling to cross a moat. In many ways, this should come as no surprise. The expression for the transmission $T$ for a particle with mass $m$ and energy $E> V_0$ across a barrier of height $V_0$ and width $b$ is
\begin{equation}
T = {1 \over 1+ {V_0^2 \over 4 E (E - V_0)} \sin^2{\left[ \sqrt{{2m b^2 \over \hbar^2}(E-V_0)}\right] } }.
\label{trans}
\end{equation}
As $V_0$ increases for a given energy, the transmission will generally decrease (resonance conditions excepted). Perhaps less appreciated by students is that a decrease in transmission also takes place (again, resonance conditions excepted) for $V_0 < 0 < E$ as the magnitude of $V_0$ increases. Equation~(\ref{trans}) applies for all (positive) energies, and the decrease in transmission for large $|V_0|$ is simply $T \propto 1/|V_0|$.

The motivation for studying such a problem arises through an interest in the physical properties of solids. A solid is composed of a (functionally) infinite periodic array of atoms. Upon the release of a valence electron, each positively charged ion is viewed as a negative potential from the valence electron's point of view. Many of a solid's features, such as its band structure and transport properties, depend primarily on the states of these valence electrons.

The issue connected to our problem comes when constructing valence wave functions - near the cores, valence wave functions have rapid oscillations and require a large number of Fourier components to reconstruct. We can make the valence wave functions easier to construct by using the orthogonalized plane-wave\cite{herring40} method and pseudopotential method,\cite{phillips59,antoncik59} which are both ways to use the bound states of a potential well to help find the scattering states much more quickly. The essence of the pseudopotential method is that the existence of these bound states resembles the presence of a repulsive potential. Note that we typically don't know the exact bound states of an atom in a periodic lattice. Fortunately, the deeply bound core states of each atom are roughly the same whether the atom is alone in a vacuum or near other atoms in a periodic array. Thus we can simplify the pseudopotential method by using bound states of a free atom to approximate the bound states of an atom in a solid.

\begin{figure}[ht]
    \centering
    \includegraphics[width=\linewidth]{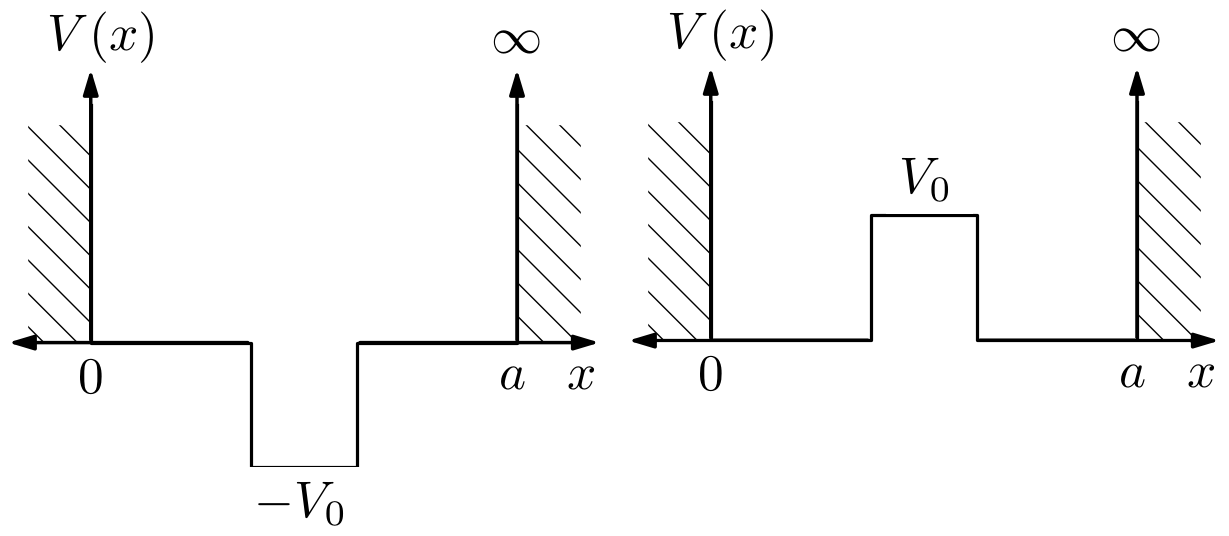}
    \caption{The general form of (a) the negative rectangular potential (moat, $ V_0 < 0 $) and (b) the positive rectangular potential (wall, $ V_0 > 0 $). The surrounding infinite square well extends between $ 0 < x < a $, and the enclosed moat and wall are both centred at $ x = a/2 $ and have a width $ b $. The goal of this paper is to show that the moat and the wall have scattering states with similar properties that make it clear that both potentials act as barriers, and therefore both systems can be described as double well potentials.}
    \label{fig:both_potentials}
\end{figure}

The purpose of this paper is to show that, in the limit of strong attractive potentials, a ``scattering'' particle will encounter an attractive potential in a manner similar to the way it encounters a repulsive potential, precisely because of the aforementioned pseudopotential effect. For simplicity we will not deal with a periodic solid, but instead focus on a single barrier, either positive or negative, that separates two regions, i.e. a double well potential. The demonstration of this method will be done using finite rectangular potential barriers centred in and contained within an infinite square well --- see Fig.~\ref{fig:both_potentials}. 
The arguments presented in this paper build on those provided in Ref.~[\onlinecite{ibrahim18}], where similar ideas were put forward but were described using simpler and idealized $ \delta $-function potentials. The finite rectangular potential is in many ways more realistic; for example, the barrier potential is now {\it not} automatically invisible to odd-parity eigenstates of the infinite square well. Nonetheless, this problem is still exactly solvable through simple analytical means. Finally, we will introduce the orthogonalized plane-wave (OPW) method and the pseudopotential method, which both take advantage of a negative potential's bound states in order to more quickly find their scattering states. We expect the ideas presented in this paper to be suitable to
senior undergraduate students; we think they will find both the conceptual message (a well can be a barrier) as well as the technical message (the 
orthogonality requirement can appear as a pseudopotential) to be interesting and enriching for their understanding of quantum mechanics.

\section{Similarity of Wave Functions in Strongly Repulsive and Attractive Rectangular Potentials}

In this section we show that the scattering eigenstates for the finite width quantum moat and the finite width quantum wall of the same potential strength become very similar looking as the potential strength increases. By ``same strength,'' we mean that, given that their widths are the same, the quantum moat's depth is the same as the quantum wall's height.

One of the key differences between the attractive $\delta$-function potential and the attractive finite width potential is that the latter can sustain more than one bound state.\cite{othman15} As we increase the depth of the attractive well, we will experience a complication that occasionally a scattering state will become a bound state at particular ``transition potentials.'' The wave function of a state that has ``just become bound'' has features very atypical to the more deeply bound states, such as a very slow decay outside of the moat region. In fact, incorporating such a bound state into the pseudopotential would be unwise, since this bound state would differ significantly from the corresponding bound state of the isolated potential. Hence there are intervals of attractive well strength (ranges of values of $ |V_0| $) near the transition potentials that will show up as difficult regimes to accurately describe with the pseudopotential method. Indeed, as the reader may have already guessed, these regimes are very connected to the regimes corresponding to perfect transmission in Eq.~(\ref{trans}).

Figure~\ref{fig:both_potentials} illustrates both the moat and wall potentials investigated in this paper. The moat potential allows both bound states (negative energy states in the moat) and scattering states (positive energy states residing above the moat), while the wall barrier system contains only scattering states. The previous statement should be taken as a working definition of bound and scattering states for this problem --- bound (scattering) states are those with negative (positive) energies.

\begin{figure}[ht]
    \centering
    
    \includegraphics[width=\linewidth]{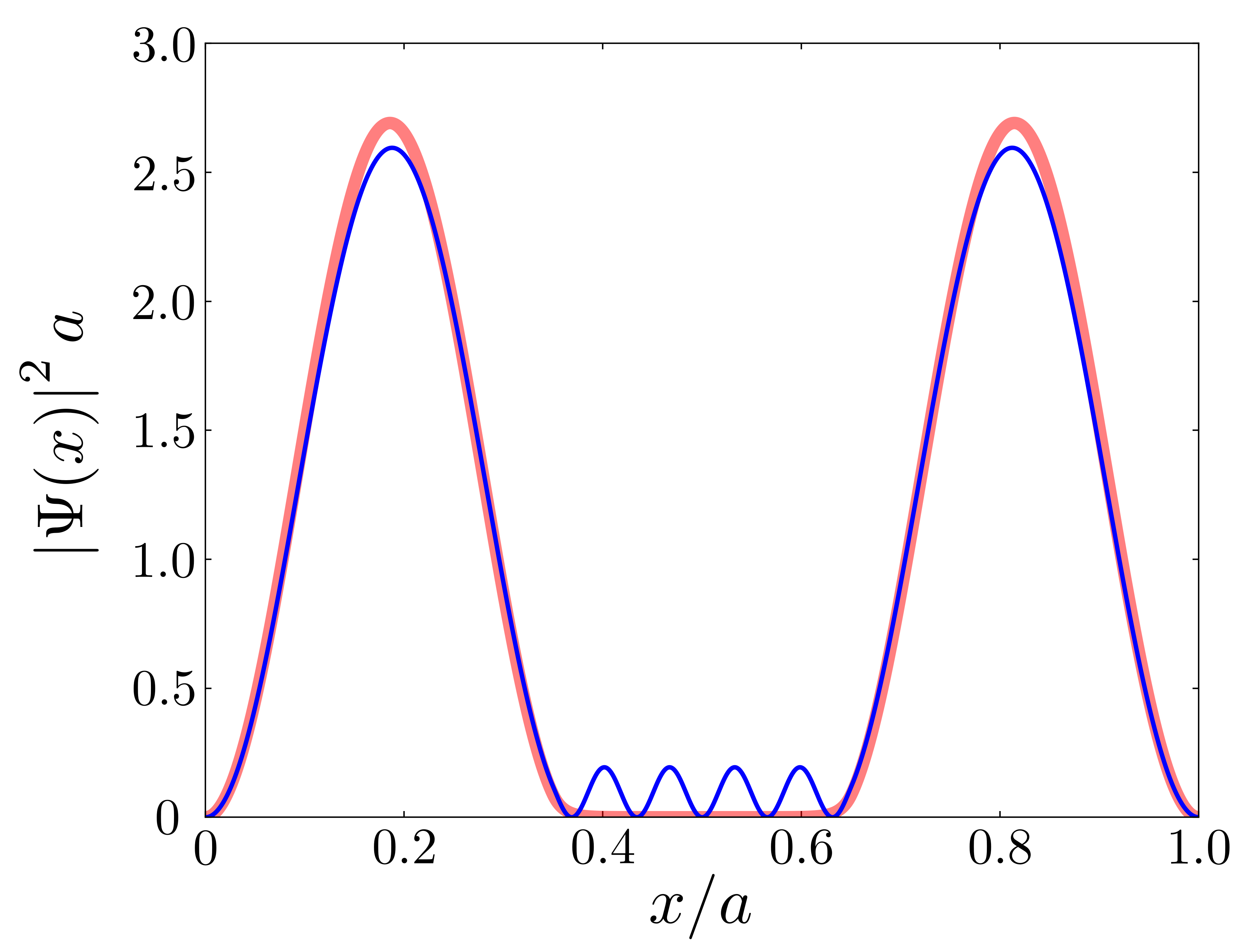}
    
    \caption{An example of a probability density for an infinite square well of width $a$ containing a rectangular negative potential barrier of width $b$ with $ V_0 = - 272E_0 $ (thin blue curve), juxtaposed with an example of a probability density for a rectangular positive potential barrier with the same width and $ V_0  = + 272 E_0 $ (thick red curve). In this case the width of both barriers is $ b/a = 0.3 $. Here we use as the unit of energy $E_0 \equiv \hbar^2/(2mw^2) $, where $w \equiv (a - b)/2$ is the width of each ``free'' region on either side of the central barrier. Both these probability densities correspond to the lowest energy scattering state of their respective potential wells. Within the barrier region, the states differ significantly: the wave function corresponding to the wall barrier is almost completely suppressed while the wave function corresponding to the moat barrier contains small oscillations. The important region to describe properly, however, is outside the barrier region where the two wave functions are very similar to one another.}
    \label{fig:similar_wave_functions}
\end{figure}

Using a barrier width of $b=0.3a$ and a large barrier strength $|V_0|$, Fig.~\ref{fig:similar_wave_functions} shows an example of two exact wave function solutions (the red, thicker curve for the positive potential wall, and the blue, thinner curve for the negative potential moat) for barriers of equal absolute strength. Details concerning the plot are present in the figure caption, and their resemblance to one another is the subject of this paper. It is important to realize that the thicker red curve represents the lowest energy state of the positive barrier potential, while the blue, thinner curve represents the fifth (5$^{th}$) excited state for this negative barrier potential, since for the parameters in this figure, there are five (5) states bound in the inner moat region.

Apart from the low-lying oscillations present in the wave function with the attractive potential barrier (i.e. moat), the two look nearly the same. In particular, both show the generic superposition of probability density in the ``free'' region, and the relative absence of probability density in the barrier region, that is so characteristic of the solution for the ground state of a double well potential with a repulsive barrier in its centre.

% Presenting and rationalizing the use of Delta A
To quantify how similar two wave functions are, we have defined the dimensionless figure of merit $ \Delta A $, given by
\begin{equation} \label{eq:delta_A}
    \Delta A \equiv \int_0^a \mathrm{d}x \left| \left| \psi_{\rm 1}(x) \right|^2 - \left| \psi_{\rm 2}(x) \right|^2 \right| ,
\end{equation}
\vskip0.1in
\noindent where $ \psi_{\rm 1} $ and $ \psi_{\rm 2} $ are the two wave functions to be compared. This definition provides a measure of the differences in probability density between the two wave functions - the more similar two probability densities are, the smaller $ \Delta A $ becomes. Its most obvious shortcoming, that it would return a large value for two similarly shaped wave functions in different locations, is not relevant for our states of interest since they are completely distributed between $ x = 0 $ and $ x = a $ and we place the wall or moat in the same central region.

% Explain the graph where Delta A changes with potential
The wave functions $ \psi_{\rm 1} $ and $ \psi_{\rm 2} $ could represent the two wave functions in Fig.~\ref{fig:similar_wave_functions}, but we are in fact mostly concerned with the similarity of two wave functions for the same model potential - the lowest even and odd scattering states for the moat potential, whose $ \Delta A $ will be denoted by $ \Delta A_{\rm moat} $, and the lowest even and odd scattering states for the wall potential, whose $ \Delta A $ will be denoted by $ \Delta A_{\rm wall} $. This is because one of the ``tell-tale'' characteristics of a particle in a double well potential is the fact that the two lowest (scattering) states have nearly identical probability densities. For example, had we drawn the probability distribution of the first excited state for the positive barrier potential in Fig.~\ref{fig:similar_wave_functions}, it would have looked nearly identical to that of the positive barrier potential's ground state (the red curve). We want to show that this is also the case for the attractive potential moat. We remind the reader that for the positive barrier potential, as the barrier height becomes very large, the even and odd scattering states become even and odd combinations of the ground state for an isolated well of width $w$. This means their probability densities become essentially identical to one another. We will see to what degree this occurs for a moat separating the two free regions.

We proceed as follows. First we picked a large range of potential strengths $ \vert V_0 \vert $, and generated walls and moats of each potential strength for this range. More informally, we created ``walls that were as tall as the moats were deep,'' and did this for many values of $ \vert V_0 \vert $. For every value of $ \vert V_0 \vert $, we calculated $ \Delta A_{\rm moat} $ and $ \Delta A_{\rm wall} $. This procedure was done for wall/moat pairs of four different widths: $ 0.20 $, $ 0.10 $, $ 0.05 $, and $ 0.02 $. The results are presented in Fig.~\ref{fig:dA_values}. The analytic solutions to all of these scattering wave functions are available in Appendix A.

%fig. 3
\begin{figure}[ht]
    \centering
    
    \includegraphics[width=\linewidth]{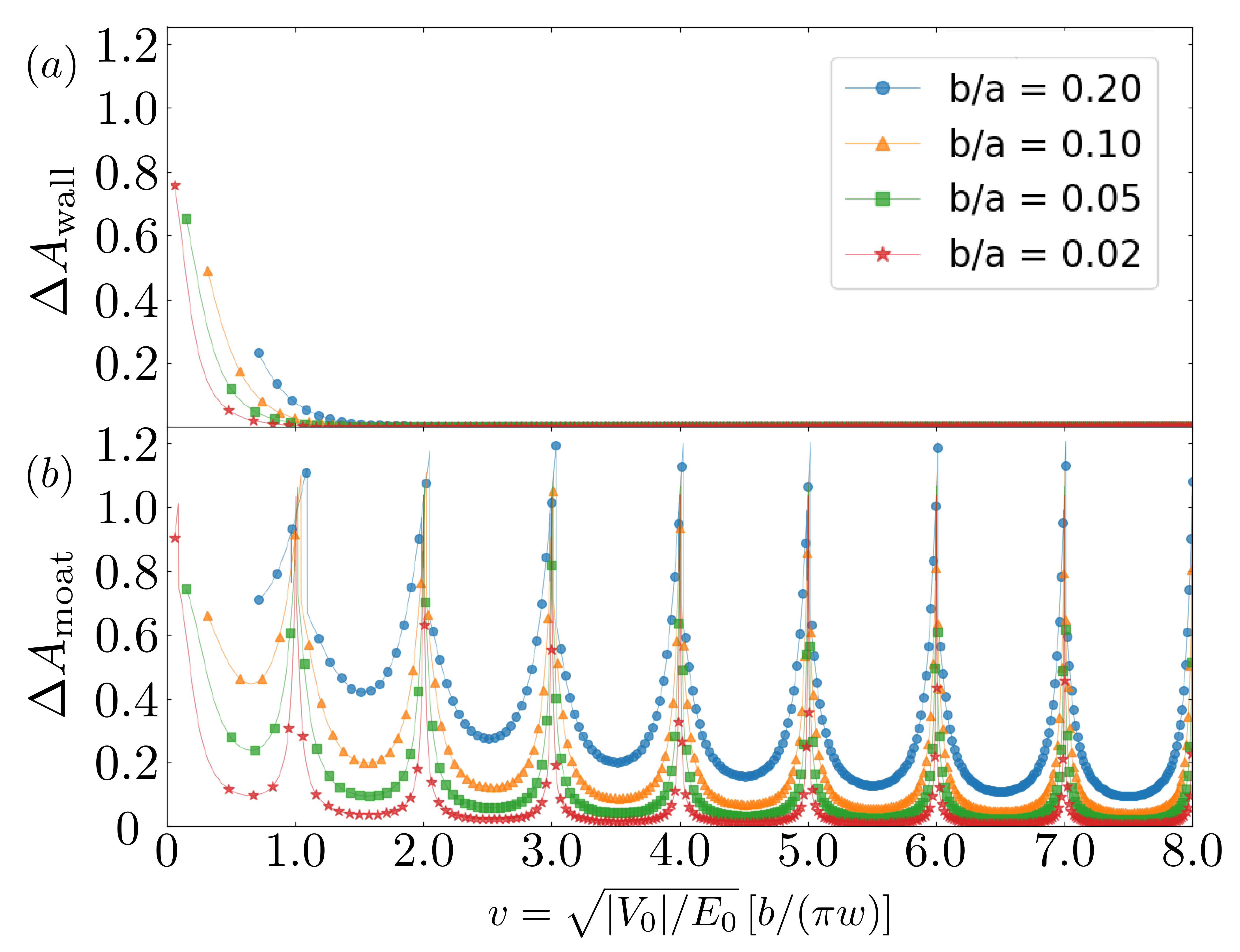}
    
    \caption{(a) $ \Delta A_{\rm wall} $  and (b) $ \Delta A_{\rm moat} $ as a function of $ v = \sqrt{ \vert V_0 \vert / E_0} \left[ b / (w \pi) \right] $. It is clear from figure (b) that something peculiar is occurring as $v$ takes on integer values. This is just the condition that another of the quantum moat's (formerly scattering) states becomes bound. It is therefore no surprise that when this is about to happen to the scattering state with the lowest energy, it will look very different from the scattering state with the second lowest energy, and $ \Delta A_{\rm moat} $  will increase sharply. In the wall barrier case (a) the measure of similarity goes quickly to zero. This shows the generic and expected behaviour for the even and odd scattering wave functions in a double well potential.}
    \label{fig:dA_values}
\end{figure}

% even wall vs odd wall
% even moat and odd moat
Figure~\ref{fig:dA_values}~(a) compares the probability densities for the lowest even and odd scattering wave functions for the wall. As expected, these become more similar ($\Delta A_{\rm wall} \rightarrow 0$) as the potential barrier heights increase. Note that for sufficiently high barriers, the actual width of the barrier becomes immaterial.

When it comes to the moat, there is a complication mentioned earlier in that, as the moat is made deeper, there exist potentials at which a scattering state transitions to a bound state. If we define a dimensionless value of the potential strength through $ v \equiv \sqrt{ \vert V_0 \vert / E_0} \left( b / w \pi \right) $, then at every even integer $ v $, an even scattering state transitions to an even bound state, and at every odd integer $ v $, an odd scattering state transitions to an odd bound state. Figure~\ref{fig:dA_values}~(b) shows how $ \Delta A_{\rm moat} $ varies with the potential strength $ v $, and indicates that the lowest even and odd scattering states of the moat are the most different at integer values of $ v $ and the most similar at half-integer values of $ v $. This means that $ \Delta A_{\rm moat} $ is a sensible measure of the similarity of the two lowest scattering states only for values of $ v $ between and well away from the integers. As expected, the wave functions are the most similar between transitions and the most different near them. When considering values of $ v $ away from the integers, there is a definite decrease of $ \Delta A_{\rm moat} $ with increasing $ v $, but it is considerably slower compared to the case of $ \Delta A_{\rm wall} $. A more quantitative analysis is provided in Appendix~B. 
 
As opposed to a wall, the width $ w $ of a moat remains an important factor in the trend for much higher moat depths. Note moreover, that as the width of the moat decreases, the range of values of $v$ over which this resonance behavior occurs decreases considerably. In fact, in the limit of $\delta$-function barriers ($b/a \rightarrow 0$), there are no resonances, and Fig.~\ref{fig:dA_values}~(b) illustrates that this is achieved by having the resonance regions become reduced in scope as the well width decreases.

% Brief summary of the section
To summarize this section it is clear that for the moat, if we focus on $ \Delta A_{\rm moat} $ for values of $v$ well away from the anomalous peaks due to scattering-bound state transitions (values of $v$ close to the minima in Fig.~\ref{fig:dA_values}~(b)), then the two wave functions (the even and odd scattering states of lowest energy) behave more like those in the quantum wall barrier problem as $|V_0|$ increases. In fact, since the occurrence of anomalous peaks in $ \Delta A_{\rm moat} $ varies as $|V_0|^2$, the range of potentials where this is true increases as $|V_0|$ increases.
%end of similarity section

\section{Pseudopotential Method}
A deeper understanding of the effectiveness of a quantum moat as a barrier can be realized by recognizing the role of the bound states as a pseudopotential. As seen from the point of view of a particle in a scattering state, the bound states give rise to an effective repulsive potential. The essence of the pseudopotential method has been explained in Ref.~[\onlinecite{ibrahim18}], so here we merely provide a brief overview of the orthogonalized plane-wave (OPW) method and how it leads to the pseudopotential method.

The difficulty with constructing wave functions using the ordinary plane wave expansion method is that a large number of Fourier components (i.e. the sine functions of our basis) is typically required to reproduce the rapid oscillations present in the core regions of a lattice. As a remedy, the OPW method takes advantage of the orthogonality requirement of solutions to the Schr\"odinger equation. Instead of using a basis of ordinary plane waves\cite{remark1} $ \ket{\phi_j} $, the OPW method uses a basis of modified plane waves, where the core state components are projected out from each plane wave basis state.

We use a new OPW basis $ \left\{ \ket{\tilde{\phi}_j} \right\} $ given by
\begin{equation} \label{eq:basis_change}
	\ket{\tilde{\phi_j}} = \ket{\phi_j} - \sum_B \ket{\phi_B} \! \braket{\phi_B}{\phi_j} ,
\end{equation}
where $ \left\{ \ket{\phi_j} \right\} $ is the ordinary plane wave basis and the sum is over all core bound states $ \left\{ \ket{\phi_B} \right\} $. The core region's rapid oscillations are present in the new basis states, making them more efficient at constructing the scattering states. The pseudopotential method is a realization of the OPW method, and allows one to formulate an effective potential using these core bound states.

To see how this comes about, we express a typical (scattering) state in terms of the OPW basis,
\begin{equation} \label{eq:scattering_state_expansion}
	\ket{\psi} = \sum_j c_j \ket{\tilde{\phi}_j}
\end{equation}
\noindent
and substitute this into the Schr\"odinger equation
\begin{equation}
	(\hat{H}_0 + \hat{V}) \ket{\psi} = E \ket{\psi} ,
\end{equation}
\noindent
where $ \hat{H}_0 $ is the kinetic energy term, $ \hat{V} $ is the one-body potential of interest, and $ E $ is the energy of the scattering state. Taking the inner product with $\bra{\tilde{\phi}_i}$, we get
\begin{equation} \label{eq:pseudo_Schrodinger}
	\sum_j \bra{\phi_i} \left[  \hat{H}_0 + \hat{V} + \sum_B (E - E_B) \ket{\phi_B} \bra{\phi_B} \right] \ket{\phi_j}c_j = E c_i ,
\end{equation}
\noindent
where $ E_B $ is the energy of each bound state. Equation~(\ref{eq:pseudo_Schrodinger}) is the Schr\"odinger matrix equation with an effective potential, usually referred to as a pseudopotential $\hat{V}_{ps}$, which is a generalized version of Eq.~(14) in Ref.~[\onlinecite{ibrahim18}],
\begin{equation} 
\label{eq:pseudopotential}
    \hat{V}_{ps} \equiv \hat{V} + \sum_B (E - E_B) \ket{\phi_B} \bra{\phi_B} .
\end{equation}

The pseudopotential method transforms the original Hamiltonian into a new type of ``pseudo-Hamiltonian''. Its solutions (the ``pseudo-solutions'' $ \ket{\psi} $) typically lack the rapid oscillations found in the solutions to the original Hamiltonian, and thus require fewer Fourier components in their construction. Most often there is no need to transform back, as the pseudo-solutions are generally accurate in the region away from the core, which is the primary region of interest when studying many of a metal's properties.

The matrix elements of Eq.~(\ref{eq:pseudo_Schrodinger}) consist of the sum of three components,
\begin{equation}
    \bra{\phi_i} \hat{H}_0 \ket{\phi_j}, \ \ \ \ \bra{\phi_i} \hat{V} \ket{\phi_j}, \ \ \ \ \mathrm{and} \ \ \ \ \sum_B (E - E_B) \braket{\phi_i}{\phi_B} \braket{\phi_B}{\phi_j}.
    \label{three_terms}
\end{equation}
Because we know the energies and wave function forms for both the basis states and bound states, every component of the above three terms can be found analytically. But recall that our motivation for studying the pseudopotential was to calculate the scattering states of a periodic array of atoms in a metal. Unfortunately, the energies and wave functions of the bound states of this atomic lattice are generally unknown. But because the deeply bound states of a free atom are nearly identical to bound states of an atom in an array, the former can be used as approximations of the latter. For our model of the problem, this corresponds to using the bound states of an isolated rectangular finite potential well of width $ b $ to approximate the bound states of a rectangular potential moat in an infinite square well. As long as the former are reasonable representations of the latter, then the scattering states should be accurately produced. All three terms in Eq.~(\ref{three_terms}) are determined analytically in Appendix B.

\begin{figure}[ht]
    \centering
    
    \includegraphics[width=\linewidth]{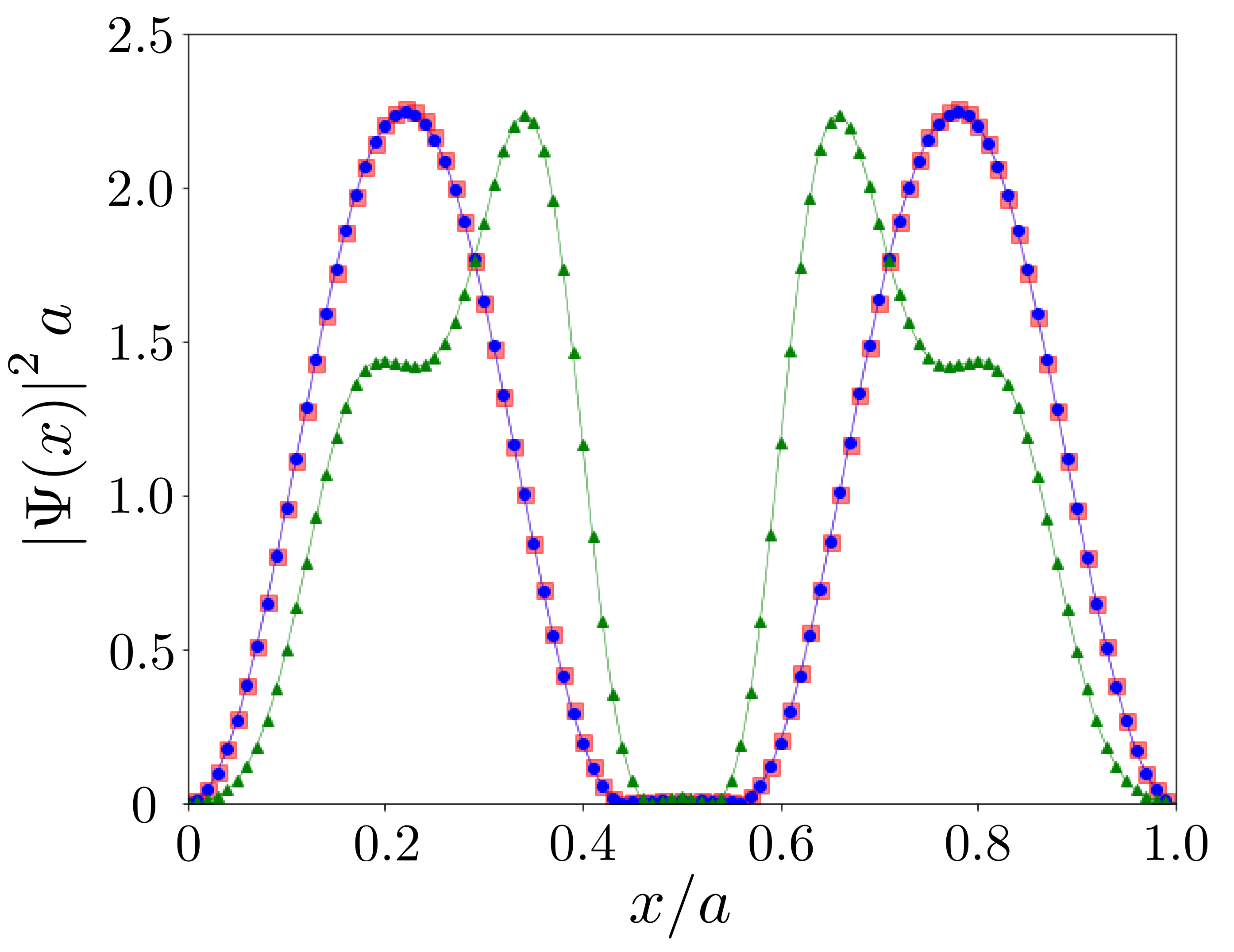}
    
    \caption{The analytic solution (red squares) to the lowest even scattering wave function of a moat of width $ b/a = 0.10 $ and a potential strength of $ V_0/E_0 = -4000 $. Ordinary plane-wave expansion (green triangles) and the pseudopotential method using approximate bound states (blue circles) were used to approximate this wave function, both using a matrix of size $ N = 10 $. The ordinary method does a poor job (it can of course be improved with an increased matrix size), while the pseudopotential method is very accurate with a small matrix representation. 
%managed to achieve $ \Delta A = 0.54 $, %0.537
%while the pseudopotential method achieved $ \Delta A = 0.0046$. %0.00458 $.
%The ordinary method can reach a similar value of $ \Delta A = 0.0057$ %0.00574 $ 
%with $ N = 57 $.
}
    \label{fig:example_wave_functions}
\end{figure}

In the case where a bound state of the atom in an array is not sufficiently tightly bound, it will not be approximated well by the corresponding bound state of a free atom. As a result, we will not include it in the pseudopotential method. Consequently, we require a criterion to determine when an approximate bound state is bound strongly enough to use in the pseudopotential method. In addition, this criterion must be independent of the exact form or energies of the actual bound states, to which we would normally not have access. For our purposes we consider a state to be deeply bound if over $ 99 \% $ of its probability density is contained in $ 0 < x < a $, or in other words, if $ \int_0^a \mathrm{d} x \vert \phi_B(x) \vert ^ 2 > 0.99 $. This ensures that the analytic expressions that we adopt for the bound states (see Appendix C) are accurate. A simple example of the improvement in efficiency is presented in Fig. \ref{fig:example_wave_functions}, where the analytic solution can be well-approximated using the pseudopotential method with a far smaller matrix compared to ordinary plane-wave expansion. This figure makes it clear that the notion of a moat as a quantum barrier applies just as well for a finite-width potential as it did for the $\delta$-function potential.\cite{ibrahim18}

%Consequently, we need a criterion to determine when an approximate bound state is bound strongly enough to use in the pseudopotential method. In addition, this criterion must be independent of the exact form or energies of the actual bound states, to which we would normally not have access.
%
%What we do know is that the bound state wave functions of the quantum moat centred in an infinite square well are completely contained in the infinite square well, i.e. between $ 0 < x < a $; their amplitude is zero elsewhere. Meanwhile, the corresponding bound state wave functions of the finite potential well (same width and depth as the moat) ``spill out'' (with exponentially small tails) of that region, but the more deeply bound they are, the more they are contained for all purposes in the region  $ 0 < x < a $. We therefore use a criterion that a bound state is considered deeply bound if over $ 99 \% $ of its probability density is contained in $ 0 < x < a $, or in other words, if $ \int_0^a \mathrm{d} x \vert \phi_B(x) \vert ^ 2 > 0.99 $.

%fig. 5

%Figure~\ref{fig:pseudopotential_example} displays two different $ \Delta A $ calculations both in the form
%\begin{equation} \label{eq:exact_vs_diag_dA}
%	\Delta A \equiv \int_0^a \mathrm{d}x \left| \left| \psi_{\rm exact}(x) \right|^2 - \left| \psi_{\rm diag}(x) \right|^2 \right| ,
%\end{equation}
%where $ \psi_{\rm exact} $ is calculated using the analytical solution to the lowest scattering state, and $ \psi_{\rm diag} $ is calculated using one of two matrix diagonalization methods -  the ordinary diagonalization method or the pseudopotential diagonalization method. All wave functions in Fig.~\ref{fig:pseudopotential_example} are even-parity. It is clear that the pseudopotential method is able to compose the lowest scattering state to a certain accuracy much more efficiently than ordinary plane-wave expansion, except at potentials where the lowest scattering state is nearly bound, where it it only marginally more efficient. 

%end of pseudopotential section

\section{Summary}
In this paper we have examined the cases of double well potentials in which the potential barrier was either a negative potential moat or a positive potential wall. Solving the Schr\"odinger equation for both these systems showed that the lowest scattering states in both potentials appeared very similar except when, in the case of the negative potential, a scattering state was on the verge of becoming a bound state. We have also shown how to apply the pseudopotential method to find the scattering states, and that this method is more efficient than the original plane wave expansion method. Most importantly, the use of a pseudopotential explains why a quantum moat is just as effective as a positive barrier in creating an effective double well potential.

We also end with a message of what this is {\it not}: it is tempting to apply classical thinking to the notion of ``using bound wave functions as a wall'' in the pseudopotential method. For example, it is reasonable to assume that the reason the moat acts like a wall is because the scattering particles cannot occupy the same positions as the bound particles trapped in the moat's centre. Or, if the particles are charged, like electrons, then one might assume that the scattering electrons are electrically repelled by the collection of bound electrons in the moat. These arguments are incorrect. We emphasize that the scattering wave functions of the moat will still behave as if the moat were a wall even if none of the moat's bound states are actually occupied, or if the potential were the result of something other than charged particles. In other words, the arguments presented in this paper depend on neither Coulomb repulsion nor the Pauli Exclusion Principle. They depend only on the requirement that the solutions to the Time-Independent Schr\"odinger equation are orthonormal to each other.

\begin{acknowledgments}

This work was supported in part by the Natural Sciences and Engineering Research Council of Canada (NSERC) and by the Department of Physics at the University of Alberta. We also wish to acknowledge a University of Alberta Teaching and Learning Enhancement Fund (TLEF) grant received a number of years ago that in fact stimulated this work. We also thank Don Page and Joseph Maciejko for discussions and interest in this problem. 

\end{acknowledgments}

\appendix

\section{Analytic Solutions of Wave Functions}

In the following sections of the Appendix, we describe the solutions and energies of finite square potential walls and moats surrounded by an infinite square well spanning $ 0 < x < a $. We refer to Region \RNum{1} as $ 0 < x < w $, Region \RNum{2} as $ w < x < w + b $, and Region \RNum{3} as $ w + b < x < a $, where $ b $ is the width of the central barrier region and $ w = (a - b)/2 $. For the bound states of the finite potential well system, Regions \RNum{1}, \RNum{2} and \RNum{3} refer to the same ranges of $ x $ values, even though the wave functions spill into $ x < 0 $ and $ x > a $.

\section*{(i) Moat - Scattering States}

For the scattering states of the quantum moat, we have defined $ k \equiv \sqrt{2mE/\hbar^2} $ and $ q \equiv \sqrt{2m(E - V_0)/\hbar^2} $.

\noindent
For even states, the acceptable values of $ k $ and $ q $ are the solutions to
\begin{equation} \label{eq:moat_eq_scat_even}
    \tan( k w ) = \frac{ k }{ q } \cot( qb/2 ).
\end{equation}

\noindent
The even wave function is
\begin{equation}
    \Psi_{moat, even, scat}(x) =
        \begin{cases}
            A_e \, \sin(kx), & \mathrm{\RNum{1}} \\
            A_e \, L_e \, \cos q \big( x - \frac{a}{2} \big), & \mathrm{\RNum{2}} \\
            -A_e \, \sin k (x - a), & \mathrm{\RNum{3}}
        \end{cases}
\end{equation}
\noindent
where the even amplitude is given by
\begin{equation} \label{eq:moat_amp_scat_even}
	\left| A_e \right|^2 = \frac{1}{w} \left[ 1 - \frac{\sin(2kw)}{2kw} + \frac{b L_e^2}{2w} \left[ 1 + \frac{\sin(qb)}{qb} \right] \right]^{-1},
\end{equation}
\noindent
and $ L_e = \sin( k w ) / \cos( qb/2 ) $. 

\noindent
For odd states, the acceptable values of $ k $ and $ q $ are the solutions to
\begin{equation} \label{eq:moat_eq_scat_odd}
    \tan( k w ) = - \frac{ k }{ q } \tan( qb/2 ).
\end{equation}

\noindent
The odd wave function is
\begin{equation}
    \Psi_{moat, odd, scat}(x) =
        \begin{cases}
            A_o \, \sin(kx), & \mathrm{\RNum{1}} \\
            A_o \, L_o \, \sin q \big( x - \frac{a}{2} \big), & \mathrm{\RNum{2}} \\
            A_o \, \sin k (x - a). & \mathrm{\RNum{3}}
        \end{cases}
\end{equation}
\noindent
where the odd amplitude is given by
\begin{equation} \label{eq:moat_amp_scat_odd}
	\left| A_o \right|^2 = \frac{1}{w} \left[ 1 - \frac{\sin(2kw)}{2kw} + \frac{b L_o^2}{2w} \left[ 1 - \frac{\sin(qb)}{qb} \right] \right]^{-1},
\end{equation}
\noindent
and $ L_o = - \sin( k w ) / \sin( qb/2 ) $.

\noindent
To find the energies and amplitudes of the scattering states of the moat, define $ z^2 =  \vert E \vert / E_0 $ and $ z_0^2 = |V_0| / E_0 $, substitute $ kw = z $, $ qw = \sqrt{z_o^2 + z^2} $, and $ qb = (b/w) \sqrt{z_o^2 + z^2} $ into Eqs.~(\ref{eq:moat_eq_scat_even}, \ref{eq:moat_eq_scat_odd}), and solve each numerically for $ z $.

\section*{(ii) Wall - Scattering States}

For the scattering states of the quantum wall, we have defined $ q \equiv \sqrt{2mE/\hbar^2} $ and $ \kappa \equiv \sqrt{2m(V_0 - E)/\hbar^2} $.

\noindent
For even states, the acceptable values of $ q $ and $ \kappa $ are the solutions to
\begin{equation} \label{eq:wall_eq_scat_even}
    \tan( q w ) = - \frac{ q }{ \kappa } \coth( \kappa b/2 ).
\end{equation}

\noindent
The even wave function is
\begin{equation}
    \Psi_{wall, even, scat}(x) =
        \begin{cases}
            A_e \, \sin( q x ), & \mathrm{\RNum{1}} \\
            A_e \, L_e \, \cosh \kappa \big( x - \frac{a}{2} \big), & \mathrm{\RNum{2}} \\
            -A_e \, \sin q (x - a). & \mathrm{\RNum{3}},
        \end{cases}
\end{equation}
\noindent
where the even amplitude is given by
\begin{equation} \label{eq:wall_amp_scat_even}
    \left| A_e \right|^2 = \frac{1}{w} \left[ 1 - \frac{\sin(2qw)}{2qw} + \frac{b L_e^2}{2w} \left[ \frac{\sinh(\kappa b)}{\kappa b} + 1 \right] \right]^{-1},
\end{equation}
\noindent
and $ L_e = \sin( q w ) / \cosh( \kappa b/2 ) $.

\noindent
For odd states, the acceptable values of $ q $ and $ \kappa $ are the solutions to
\begin{equation} \label{eq:wall_eq_scat_odd}
    \tan( q w ) = - \frac{ q }{ \kappa } \tanh( \kappa b/2 ).
\end{equation}

\noindent
The odd wave function is
\begin{equation}
    \Psi_{wall, odd, scat}(x) =
        \begin{cases}
            A_o \, \sin( q x ), & \mathrm{\RNum{1}} \\
            A_o \, L_o \, \sinh \kappa \big( x - \frac{a}{2} \big), & \mathrm{\RNum{2}} \\
            A_o \, \sin q (x - a). & \mathrm{\RNum{3}}
        \end{cases}
\end{equation}
\noindent
where the odd amplitude is given by
\begin{equation} \label{eq:wall_amp_scat_odd}
    \left| A_o \right|^2 = \frac{1}{w} \left[ 1 - \frac{\sin(2qw)}{2qw} + \frac{b L_o^2}{2w} \left[ \frac{\sinh(\kappa b)}{\kappa b} - 1 \right] \right]^{-1},
\end{equation}
\noindent
and $ L_o = - \sin( q w ) / \sinh( \kappa b/2 ) $.

\noindent
To find the energies and amplitudes of the scattering states of the wall, define $ z^2 =  \vert E \vert / E_0 $ and $ z_0^2 = |V_0| / E_0 $, substitute $ qw = z $, $ \kappa w = \sqrt{z_o^2 - z^2} $, and $ \kappa b = (b/w) \sqrt{z_o^2 - z^2} $ into Eqs.~(\ref{eq:wall_eq_scat_even}, \ref{eq:wall_eq_scat_odd}), and solve each numerically for $ z $.

\section*{(iii) Bound States of the Finite Potential Well}

We {\it could} solve for the bound states for a finite potential well contained within an infinite square well, but this exercise would be contrary to the philosophy behind the pseudopotential method. Instead, we solve for the bound states of the finite potential well in free space, given by
\begin{equation}
    V(x) =
        \begin{cases}
            V_0, & -b/2 < x < b/2 \\
            0, & \mathrm{elsewhere.}
        \end{cases}
\end{equation}

We define $ q \equiv \sqrt{2m(E_B - V_0)/\hbar^2} $ and $ \kappa \equiv \sqrt{-2mE_B/\hbar^2} $. Then, for even states, the bound state energies satisfy
\begin{equation} \label{eq:finite_eq_even}
    \kappa = q \tan( qb/2 ).
\end{equation}

\noindent
The even wave function is ($x^\prime \equiv x - a/2$)
\begin{equation}
    \Psi_{finite, even}(x) =
        \begin{cases}
            A_e \, \exp( \kappa (x^\prime + {b/2}) ), & \mathrm{\RNum{1}} \\
            A_e \,   {\cos \big( q  x^\prime \big) \over \cos ( {q b /2} )}, & \mathrm{\RNum{2}} \\
            A_e \, \exp( - \kappa (x^\prime  - {b/2}) ), & \mathrm{\RNum{3}}
        \end{cases}
\end{equation}
\noindent
and the even amplitude is given by
\begin{equation} \label{eq:finite_moat_even_amp}
    \left| A_e \right|^2 = \frac{2}{b} \ \left[{2 \over \kappa b} + {1 + \sin{(qb)}/(qb) \over \cos^2{(qb/2)}} \right]^{-1}.
\end{equation}
\noindent

For odd states, the bound state energies satisfy
\begin{equation} \label{eq:finite_eq_odd}
    \kappa = - q \cot( qb/2 ).
\end{equation}

\noindent
The odd wave function is
\begin{equation}
 \Psi_{finite, odd}(x) =
      \begin{cases}
           \phantom{+} A_o \, \exp( \kappa (x^\prime + b/2) ), & \mathrm{\RNum{1}} \\
            -A_o \, {\sin \big(q x^\prime \big) \over \sin \big(qb/2 \big)}, & \mathrm{\RNum{2}} \\
            -A_o \, \exp( - \kappa (x^\prime - b/2)). & \mathrm{\RNum{3}}
        \end{cases}
\end{equation}
where the odd amplitude is given by
\begin{equation} \label{eq:finite_moat_odd_amp}
    \left| A_o \right|^2 = \frac{2}{b} \ \left[{2 \over \kappa b} + {1 - \sin{(qb)}/(qb) \over \sin^2{(qb/2)}} \right]^{-1}.
\end{equation}

\noindent
If we define $\tilde{z} \equiv qb/2$, and $\tilde{z}_0 = (b/2) \sqrt{-2 m V_0/\hbar^2}$, then $\kappa b/2 \equiv \sqrt{\tilde{z}_0^2 - \tilde{z}^2}$, and one can readily solve (graphically and iteratively) Eqs.~(\ref{eq:finite_eq_even}, \ref{eq:finite_eq_odd}) for $\tilde{z}$ and therefore $E_B$.

\section{Details of the Resonance Behavior in $\Delta A_{\rm moat}$}

As mentioned in the text, for $ \Delta A_{\rm wall} $, there is a sharp monotonic decrease as a function of the barrier potential strength, $v \equiv \sqrt{ \vert V_0 \vert / E_0} \left( b / w \pi \right) $. In fact, the changes in $ \Delta A_{\rm wall} $ as a function of $ v $ can be expressing using the following fit,
\begin{equation} \label{eq:dA_wall}
  \Delta A_{\rm wall} = c_1 \frac{b}{w} \exp( - c_2 v ) \quad\quad c_1 = 2.36, c_2 = 3.44 ,
\end{equation}
i.e. rapid exponential decay. The situation for $ \Delta A_{\rm moat} $ is much more complicated, but the general trend can be discerned in a fit where we only consider half-integer values of $ v $, because these are where $ \Delta A_{\rm moat} $ displays local minima and we avoid the complicated (and not representative) regions where $v$ is close to an integer value. In this case, the fit is given by
\begin{equation} \label{eq:dA_moat}
    \Delta A_{\rm moat} = \frac{b}{w} \frac{c_3}{v} \ \ \ \ \ \ \ \ c_3 = 1.39, \ \ \ \ v = \frac{1}{2}, \frac{3}{2}, \frac{5}{2}, ...
\end{equation}
which shows a much slower decrease with increasing $v$.

%fig. 4
\begin{figure}[ht]
    \centering
    
    \includegraphics[width=\linewidth]{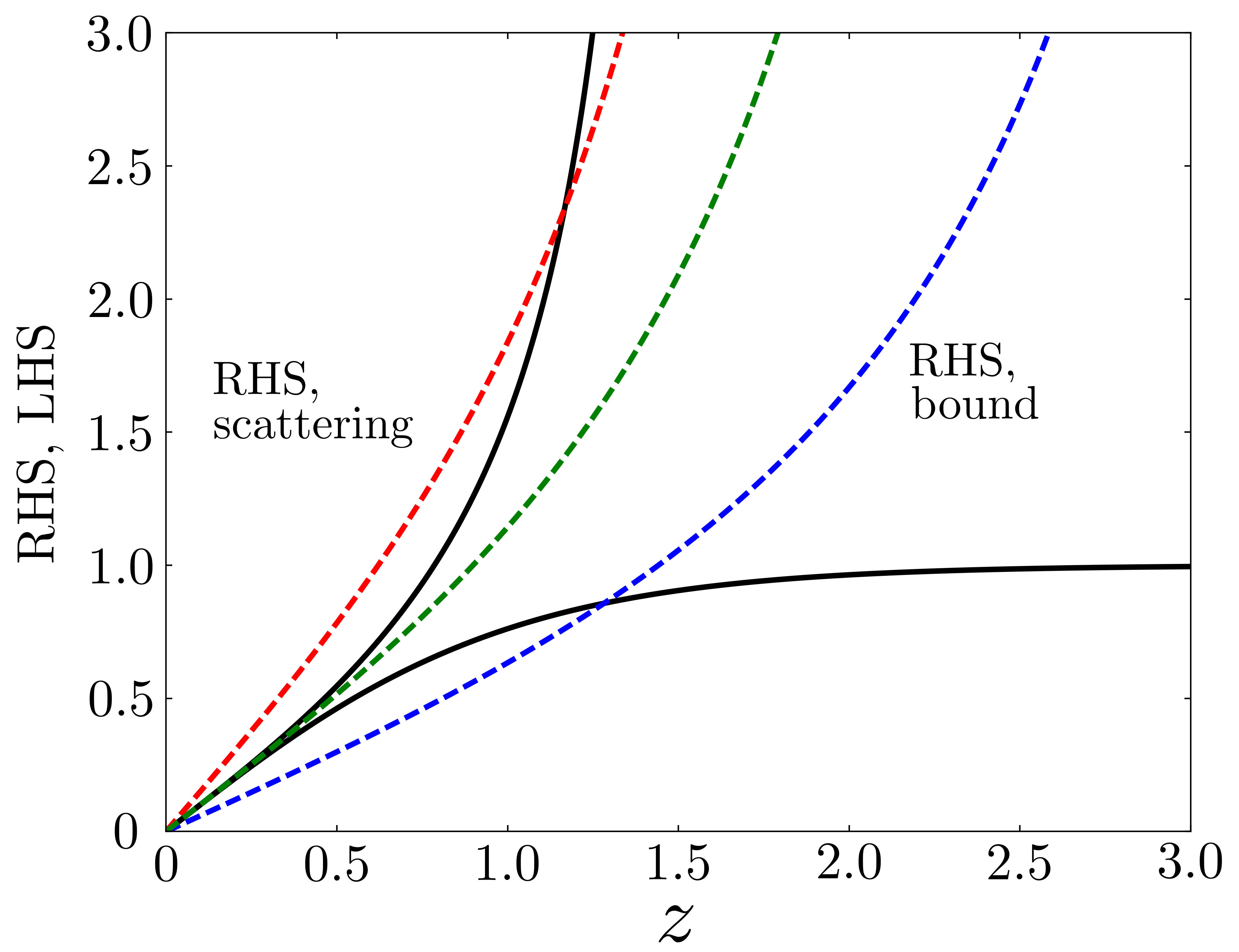}
    
    \caption{The Left-Hand-Side (LHS) of Eq. (\ref{eq:bound_eigen_z}) and of Eq. (\ref{eq:scat_eigen_z}) are both given by solid black curves, while their corresponding Right-Hand-Sides (RHSs)  are given by dashed lines. Both are plotted against 
$ z \equiv \sqrt{|E|/E_0}$. The RHSs are functions of $ z_0 \equiv \sqrt{|V_0|/E_0}$. Three separate examples of RHS equations are given. The leftmost (red dashed) curve comes from Eq. (\ref{eq:scat_eigen_z}), and occurs for a value of $ z_0 $ slightly lower than $ z_c $, corresponding to a scattering state. The rightmost (blue dashed) curve comes from Eq. (\ref{eq:bound_eigen_z}), and occurs for a value of $ z_0 $ slightly greater than $ z_c $, corresponding to a bound state. The centre (green dashed) curve is for $ z_0 = z_c $, where both RHS equations are equal, and corresponds to the transition between scattering and bound states.}
    \label{fig:eigenvalue_functions}
\end{figure}

% explaining what the peaks mean and why they occur where they do
A more quantitative understanding of the occurrence of peaks in $\Delta A_{\rm moat}$ at integer values of $v$ can be attained as follows. Using the definitions  $ k \equiv \sqrt{2mE/\hbar^2} $ (for $E>0$), $ q \equiv \sqrt{2m(E - V_0)/\hbar^2} $, and $ \kappa \equiv \sqrt{-2mE/\hbar^2} $ (for $E<0$), one can readily derive equations that determine bound state energies (see Appendix A). The bound state energies of the moat potential are determined by
\begin{equation} 
\label{eq:bound_eigen}
    \tanh ( \kappa w ) = - \frac{\kappa}{q} \tan \bigg( \frac{1}{2} q b - C \bigg),
\end{equation}
\noindent
whereas the scattering state energies of the moat potential are determined by
\begin{equation} 
\label{eq:scat_eigen}
    \tan ( k w ) = - \frac{k}{q} \tan \bigg( \frac{1}{2} q b - C \bigg),
\end{equation}
\noindent
where $ C = \pi / 2 $ for the even solution and $ C = 0 $ for the odd solution. For a graphical analysis we can make the substitutions $ z^2 =  \vert E \vert / E_0 $ and $ z_0^2 = |V_0| / E_0 $, which allows us to rewrite Eq.~(\ref{eq:bound_eigen}) as
\begin{equation} 
\label{eq:bound_eigen_z}
    \tanh ( z )  = - \frac{ z }{ \sqrt{ z_0^2 - z^2 } } \tan \bigg( \frac{ b }{ 2 w } \sqrt{ z_0^2 - z^2 } - C \bigg)
\end{equation}
\noindent
and Eq.~(\ref{eq:scat_eigen}) as
\begin{equation} 
\label{eq:scat_eigen_z}
    \tan ( z )  = - \frac{ z }{ \sqrt{ z^2 + z_0^2 } } \tan \bigg( \frac{ b }{ 2 w } \sqrt{ z^2 + z_0^2 } - C \bigg).
\end{equation}

% explaining what the peaks mean and why they occur where they do
%A more quantitative understanding of the occurrence of peaks in $\Delta A_{\rm moat}$ at integer values of $v$ can be attained as follows. Using the definitions  $ k \equiv \sqrt{2mE/\hbar^2} $ (for $E>0$), $ q \equiv \sqrt{2m(E - V_0)/\hbar^2} $, and $ \kappa \equiv \sqrt{-2mE/\hbar^2} $ (for $E<0$), one can readily derive equations that determine bound state energies (see Appendix A) [with $ C = \pi / 2 $ for the even solution and $ C = 0 $ for the odd solution]:

%\begin{equation} 
%\label{eq:bound_eigen}
%    \tanh ( \kappa w ) = - \frac{\kappa}{q} \tan \bigg( \frac{1}{2} q b - C \bigg) %\phantom{aaaaaaa}{\rm bound \ \ states}
%\end{equation}

%\begin{equation} 
%\label{eq:scat_eigen}
%    \tan ( k w ) = - \frac{k}{q} \tan \bigg( \frac{1}{2} q b - C \bigg) \phantom{aaaaaaa}{\rm scattering \ \ states}
%\end{equation}
%\vskip0.1in
%\noindent For a graphical analysis we can make the substitutions $ z^2 =  \vert E \vert / E_0 $ and $ z_0^2 = |V_0| / E_0 $, which allows us to rewrite Eqs. (\ref{eq:bound_eigen}, \ref{eq:scat_eigen}) as
%\begin{equation} 
%\label{eq:bound_eigen_z}
%    \tanh ( z )  = - \frac{ z }{ \sqrt{ z_0^2 - z^2 } } \tan \bigg( \frac{ b }{ 2 w } \sqrt{ z_0^2 - z^2 } - C \bigg) \ \ \ \ \ {\rm bound \ \  states}
%\end{equation}

%\begin{equation} 
%\label{eq:scat_eigen_z}
%    \tan ( z )  = - \frac{ z }{ \sqrt{ z^2 + z_0^2 } } \tan \bigg( \frac{ b }{ 2 w } \sqrt{ z^2 + z_0^2 } - C \bigg), \ \ \ \ \ {\rm scattering \ \  states}
%\end{equation}
%\vskip0.1in

\noindent
Figure \ref{fig:eigenvalue_functions} shows a graph of the left-hand-side (LHS) and right-hand-side (RHS) of Eqs.~(\ref{eq:bound_eigen_z}, \ref{eq:scat_eigen_z}) as a function of $ z $, from which solutions can be obtained. Details are provided in the figure caption. The transition from scattering state to bound state occurs when the slopes of the two curves are equal to one another for $z=0$. Enforcing this condition leads to the transcendental equation for the special value of the potential where this occurs, denoted by $z_c$,
\begin{equation}
\label{eq:transcend_z}
    z_c = -\tan \left( \frac{b}{2w} z_c - C \right) .
\end{equation}
\noindent
We use $ z_c^2 \equiv -2mw^2 V_c/\hbar^2 $ to define the transition potentials $ V_c $ ($ < 0 $). For large values of $ z_c $, the left and right sides of Eq. (\ref{eq:transcend_z}) are in agreement very near the asymptotes of the tangent function.  A straightforward analysis shows that for $ C=0 $, at large values of $|V_c|$, this occurs when $z_c = \pi w (2n-1)/b$, i.e. at odd integer multiples of $\pi w/b$. A similar analysis for the even solutions ($C = \pi/2$) shows that the transition from scattering to bound state occurs at even integer multiples of $\pi w/b$. Hence we achieve an understanding of the factors defining $v$ in Fig.~\ref{fig:dA_values} and why the large values of $\Delta A_{\rm moat}$ occur at integer values of $v$.

\section{Matrix Elements of the Pseudopotential Method}
Every element in the matrix of the ``pseudo-Hamiltonian'' is given by the sum of the three terms

\[
    \bra{\phi_m} \hat{H}_0 \ket{\phi_n}, \ \ \ \ \bra{\phi_m} V \ket{\phi_n}, \ \ \ \ \mathrm{and} \ \ \ \ \sum_B (E - E_B) \braket{\phi_m}{\phi_B} \braket{\phi_B}{\phi_n}.
\]

Defining $ E_n^{(0)} = n^2 \pi^2 \hbar^2 / (2 m a^2) $ to be the energy of the $ n^{th} $ state of the infinite square well ($\braket{x}{\phi_n} = A_n \sin(n \pi x / a)$, where $A_n = \sqrt{2/a}$), the first is simply given by

\[
    \bra{\phi_m} \hat{H}_0 \ket{\phi_n} = E_n^{(0)} \delta_{mn}.
\]

The second term is similar to the first but the potential is nonzero only between $ x = w $ and $ x = w + b $, so we cannot take full advantage of the orthonormality of the basis states.

\[
    V_0 \int_{w}^{w+b} A_n \sin(n \pi x/a) A_m \sin(m \pi x/a) \mathrm{dx} =
\]
\[
\begin{cases}
    \frac{ V_0 }{ 2 \pi n } \left[ \frac{ 2 \pi n }{ a } x  - \sin \left( \frac{ 2 \pi n }{ a } x \right) \right] \bigg|_{w}^{w + b}, & m = n \\
    \frac{ V_0 }{ \pi } \left[ \frac{ \sin((m - n) \pi x / a) }{ m - n }  - \frac{\sin((m + n) \pi x / a) }{ m + n } \right] \bigg|_{w}^{w + b}, & m \ne n.
\end{cases}
\]
The third is the most complicated. The energy of each bound state is given by $ E_B = - \hbar^2 \kappa^2 / 2 m $, as $ \kappa $ is defined in Appendix A - Bound States of the Finite Potential Well. The integrals given by $ \braket{\phi_m}{\phi_B} $ and $ \braket{\phi_B}{\phi_n} $ have four outcomes that depend on the orientations (even or odd) of the basis and bound states involved. If the bound state is even and the basis state is odd, or vice-versa, then the inner product is simply 0. If the basis and bound states are both even or both odd, the integrals are more complicated.

Define the definite integrals $ R_{\rm I} $, $ R_{\rm II, e} $, and $ R_{\rm II, o} $. The first is the piecewise integral over the ``free'' regions where $ V = 0 $, and due to symmetry, $ R_{\rm I} $ is the same for Regions \RNum{1} and \RNum{3}, for both even and odd bound states. The second and third integrals are piecewise integrals over the centre moat region for the even and odd bound states, respectively. For the integral $ R_{\rm I} $, $ A_{eo} $ should be replaced by with $ A_e $ when the bound state is even and $ A_o $ when the bound state is odd, as given in Appendix A - Bound States of the Finite Potential Well. Recall that in all these cases, the bound and basis states have the same orientation. Note that the solutions for the finite potential well were derived for a well of width $ b $ centred at $ x = 0 $. To align the finite potential well with the moat, use $ x^\prime = x - a/2 $.

\begin{align*}
    R_{\rm I} &= \int_0^w \mathrm{dx} \, A_{eo} \exp(\kappa (x - w)) \, A_n \sin(n \pi x / a) \\
              &= A_{eo} \sqrt{2a} \frac{\left[ \kappa a \sin(n\pi w/a) - n\pi \cos(n\pi w/a) + \exp(- \kappa w) n\pi \right]}{(\kappa a)^2 + (n\pi)^2} \\
              \\
    R_{\rm II, e} &= \int_{w}^{w + b} \mathrm{dx} \, \frac{A_e}{\cos(qb/2)} \cos(q (x - a/2)) \, A_n \sin(n \pi x / a) \\
                &= - \frac{A_e}{\cos(qb/2)} \sqrt{\frac{a}{2}} \left[ \frac{ \cos ( qa/2 - (n \pi / a + q ) x ) }{ n \pi + qa } + \frac{ \cos ( qa/2 + ( n \pi / a - q ) x ) }{ n \pi - qa } \right] \Bigg|_{w}^{w + b} \\
                \\
    R_{\rm II, o} &= \int_{w}^{w + b} \mathrm{dx} \, \frac{A_o}{\sin(qb/2)} \sin(q (x - a/2)) \, A_n \sin(n \pi x / a) \\
                &= \frac{A_o}{\sin(qb/2)} \sqrt{\frac{a}{2}} \left[ \frac{ \sin ( qa/2 - (n \pi / a + q ) x ) }{ n \pi + qa } + \frac{ \sin ( qa/2 + ( n \pi / a - q ) x ) }{ n \pi - qa } \right] \Bigg|_{w}^{w + b} \\
\end{align*}

Then the value of the integral $ \braket{\phi_B}{\phi_n} $ is given by
\begin{equation} \label{eq:pseudopotential_integral}
    \braket{\phi_B}{\phi_n} = 
    \begin{cases}
        0, & \text{One is even, the other is odd} \\
        2 R_{\rm I} + R_{\rm II, e}, & \text{Basis and bound states are both even} \\
        2 R_{\rm I} + R_{\rm II, o}, & \text{Basis and bound states are both odd.}
    \end{cases}
\end{equation}


\begin{thebibliography}{10}
\bibitem{remark1} We will continue to use the terms ``plane wave'' basis and ``plane wave'' states, but in our problem we are imbedding the potential of interest (a repulsive or attractive square well) in an infinite square well, so, we use the linear combination of plane waves which are the suitably normalized ${\rm sin}(j \pi x/a)$, $j=1, 2, 3....$, 
where ``a'' is the width of the infinite square well.
\end{thebibliography}
\end{document}